\documentclass[aps,twocolumn,prl,superscriptaddress]{revtex4-1}

\usepackage{amsmath}
\usepackage{amssymb}
\usepackage{color}
\usepackage{hyperref}
\usepackage{xcolor} 
\usepackage{soul}

\usepackage{graphicx}

\DeclareMathAlphabet{\mathitbf}{OML}{cmm}{b}{it}

\newcommand{\dbar}{{\,\mathchar'26\mkern-12mu d}}

\newcommand{\sFrac}[2]{{\textstyle\frac{#1}{#2}}}

\newcommand{\quadCdot}{\stackrel{\mbox{\small :}}{:}}

\begin{document}
\title{Statistics and properties of low-frequency vibrational modes in structural glasses}
\author{Edan Lerner}
\affiliation{Institute for Theoretical Physics, University of Amsterdam, Science Park 904, 1098 XH Amsterdam, The Netherlands}
\author{Gustavo D\"uring}
\affiliation{Facultad de F\'isica, Pontificia Universidad Cat\'olica de Chile, Casilla 306, Santiago, Chile}
\author{Eran Bouchbinder}
\affiliation{Chemical Physics Department, Weizmann Institute of Science, Rehovot 7610001, Israel}

\begin{abstract}
Low-frequency vibrational modes play a central role in determining various basic properties of glasses, yet their statistical and mechanical properties are not fully understood. Using extensive numerical simulations of several model glasses in three dimensions, we show that in systems of linear size $L$ sufficiently smaller than a crossover size $L_{D}$, the low-frequency tail of the density of states follows $D(\omega)\!\sim\!\omega^4$ up to the vicinity of the lowest Goldstone mode frequency. We find that the sample-to-sample statistics of the minimal vibrational frequency in systems of size $L\!<\!L_D$ is Weibullian, with scaling exponents in excellent agreement with the $\omega^4$ law. We further show that the lowest frequency modes are spatially quasi-localized, and that their localization and associated quartic anharmonicity are largely frequency-independent. The effect of preparation protocols on the low-frequency modes is elucidated and a number of glassy lengthscales are briefly discussed.



\end{abstract}

\maketitle

\emph{Introduction.--} Many basic mechanical, static, dynamic and thermodynamic properties of disordered systems depend on the abundance of ``soft excitations'' emerging from their intrinsic disordered nature. For example, nonlinear localized two-level systems are believed to be responsible for the anomalous thermodynamic properties of glasses at very low temperatures~\cite{Anderson,Phillips}. Plastic flow in glassy materials occurs via the collective dynamics of shear transformation zones which originate from destabilizing quasi-localized soft modes~\cite{lemaitre2004,manning2011,luka}. Relaxation processes in deeply supercooled liquids were observed to be highly correlated in space with quasi-localized soft modes~\cite{widmer2008irreversible}. Energy and heat transport~\cite{energy_transport_jamming,jamming_transport_Vincenzo}, macroscopic elasticity~\cite{ohern2003,matthieu_thesis} and sound attenuation \cite{eric_boson_peak_emt} in soft solids were all shown to depend on the density of low-lying soft modes. Thermal energy has been shown to focus spatially where localized soft modes reside~\cite{eran_thermal_fluctuations}. A first principles understanding of the abundance of such excitations is therefore of key importance.

On large enough lengthscales a glass behaves like a continuum elastic solid~\cite{barrat_3d,breakdown}, for which the lowest frequency excitations are Goldstone modes (plane waves) \cite{goldstone}. The density per unit volume of Goldstone modes is known to follow Debye's theory, $D(\omega)\!\sim\!\omega^{\dbar-1}$, with $\dbar$ being the spatial dimension and $\omega$ the mode frequency~\cite{debye}. In generic glassy systems the Goldstone modes overwhelm the density of states at low frequencies. This, in turn, poses serious difficulties in using conventional approaches to study the distribution of low-frequency glassy modes which emerge due to microscale disorder~\cite{karmakar_lengthscale}.

The jamming scenario in soft athermal glasses~\cite{ohern2003,van_hecke} or thermal hard-sphere glasses~\cite{hardSphereGlass} provides a useful theoretical framework for understanding the density of low-frequency excitations in a subclass of disordered solids in which the effective number of interactions between the constituent degrees of freedom approaches $N\dbar$ from above, with $N$ the number of particles. In particular, effective medium~\cite{wyart_emt,eric_boson_peak_emt,eric_hard_spheres_emt} and infinite-dimension replica \cite{parisi_fractal,Zamponi,silvio} calculations predict $D(\omega)\!\sim\!\omega^2$, independently of spatial dimension. Recent numerical simulations showed that this relation holds close to the jamming point, but breaks down away from it~\cite{charbonneauUniversal}.

What happens away from the jamming point, in generic glassy systems? Several theories predicted the density of {\em non-Goldstone} low-frequency modes for generic glasses, i.e. away from the jamming point, to rise from zero as $D(\omega)\!\sim\! \omega^4$~\cite{soft_potential_model,chalker,Gurevich2003,Gurevich2007}. In a recent numerical investigation of the Heisenberg spin glass model in 3D it was found that upon introducing a field which suppresses Goldstone modes, the density of states followed the $\omega^4$ law at low frequencies~\cite{parisi_spin_glass}. However, to the best of our knowledge, no such evidence has ever been presented for generic structural glasses.

\begin{figure*}[!ht]
\centering
\includegraphics[width = 0.95\textwidth]{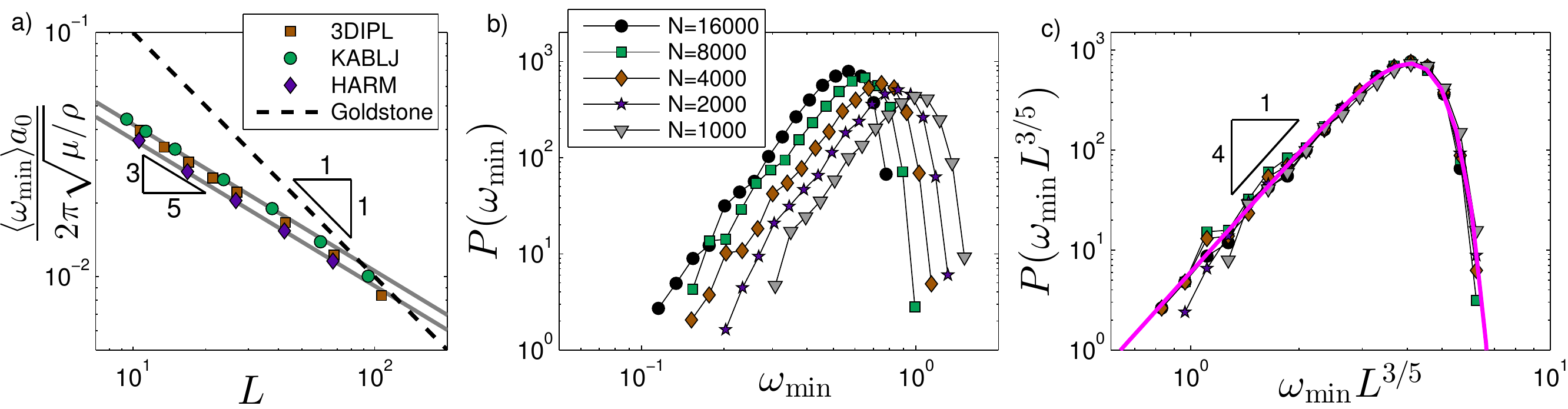}
\caption{\footnotesize (a) Sample-to-sample mean MVF $\langle \omega_{\rm min}\rangle$ rescaled by $2\pi\sqrt{\mu/\rho}/a_0$ vs.~sample length $L$, for the 3DIPL, KABLJ, and HARM models. The dashed line represents the expectation for the lowest frequency Goldstone modes $2\pi\sqrt{\mu/\rho}/L$. (b) Sample-to-sample distributions of the minimal vibrational frequency $P(\omega_{\rm min})$ for the 3DIPL system. (c) The same distributions, plotted as a function of the rescaled frequency $\omega L^{3/5}$. The continuous magenta line represents the Weibull distribution $W(y)\!\propto\!y^4e^{-(y/y_0)^5}$, with $y_0\!\approx\!4$.}
\label{MVF_fig}
\end{figure*}

In this Letter, we employ extensive numerical simulations to investigate the low-frequency vibrational modes of computer-generated structural glasses in three dimensions (3D). We show that when carefully tuning the system size $L$ to be sufficiently smaller than a crossover size $L_D$, Goldstone modes are pushed to high frequencies, revealing a density of glassy modes that follows $D(\omega)\!\sim\!\omega^4$. This result, which to the best of our knowledge is the first of its kind, is demonstrated for several popular model glasses.

Further support for this key result is presented by studying the sample-to-sample statistics of minimal vibrational frequencies (MVFs), shown to be Weibullian, with scaling exponents perfectly consistent with the $\omega^4$ law. We also study the localization and anharmonic properties of the lowest frequency modes, showing that the softest non-Goldstone modes are quasi-localized, and their associated anharmonicity and degree of localization are largely uncorrelated with their frequencies. We further examine the effect of preparation protocol on our findings, providing evidence that the $\omega^4$ law persists in glasses that are slowly cooled through the computer glass transition. Finally, we identify several lengthscales that play important roles in determining the statistics of MVFs, and briefly discuss their dependence of the glasses' preparation protocol.


\emph{Models and methods.--} We employed $3$ different computer glass-forming models in 3D: $(i)$ a binary system of soft-spheres interacting via a one-sided harmonic potential (HARM) under fixed pressure; $(ii)$ the canonical Kob-Andersen binary Lennard-Jones (KABLJ) system \cite{kablj}; $(iii)$ a binary system of point-like particles interacting via inverse power-law purely repulsive pairwise potentials (3DIPL). A complete and detailed description of the models and the numerical methods used in this work is provided in the Appendix. Unless stated otherwise, data are shown for the 3DIPL system. The ensemble of solids at zero temperature was created by a short equilibration run of each system in the liquid phase, followed by a rapid quench to zero temperature. For most system sizes the ensembles consist of a few thousand solids; for system sizes on the order of millions of particles we created a few tens or hundreds of solids.


\emph{Results.--} We begin with discussing the effects of system size on the sample-to-sample statistics of minimal vibrational frequencies (MVFs) in structural glasses. Let us assume that in the absence of Goldstone modes the low-frequency glassy modes are quasi-localized and only weakly correlated. If their frequencies are distributed according to $D(\omega)\!\sim\!\omega^\theta$ ($\theta\! > \!\dbar-1$), then a conventional scaling argument implies that the sample-to-sample mean MVF $\langle\omega_{\rm min}\rangle$ satisfies
\begin{equation}
\label{meanMVF}
\int\nolimits_0^{\langle\omega_{\rm min}\rangle}D(\omega)\,d\omega \sim N^{-1} \ \ \Longrightarrow \ \ \langle\omega_{\rm min}\rangle \sim L^{\mbox{\small$-\frac{\dbar}{1+\theta}$}}\,.
\end{equation}
The distribution $P(\omega_{\rm min}; L)$ of MVFs for different system sizes in 3D, i.e. $\dbar\!=\!3$, is expected to follow \cite{weibull1939statistical}
\begin{equation}\label{mvf_dist}
P(\omega_{\rm min}; L) = W(\omega_{\rm min}L^{\frac{3}{1+\theta}})\,,
\end{equation}
where $W(y)\!=\!\sFrac{\theta+1}{y_0^{\theta+1}}\, y^\theta e^{-(y/y_0)^{\theta+1}}$ is the Weibull distribution, and $y_0$ a scale to be discussed below. The important point is that since the lowest Goldstone frequency scales as $L^{-1}$, a crossover length $L_D$ is expected to separate the glassy $L^{-3/(1+\theta)}$ scaling and the Goldstone $L^{-1}$ scaling of MVF.

The predictions of Eqs.~(\ref{meanMVF})-(\ref{mvf_dist}) were tested by a large ensemble of glassy samples of various sizes, for the $3$ aforementioned glass-forming models. After quenching each sample, the lowest non-zero eigenvalue of the dynamical matrix ${\cal M}_{ij}\equiv \frac{\partial^2U}{\partial\vec{x}_i\partial\vec{x}_j}$, with $U$ denoting the potential energy and $\vec{x}_i$ the coordinate vector of the $i^{\mbox{\tiny th}}$ particle, was calculated. The MVF $\omega_{\rm min}$ of each sample is given by the square root of the lowest nonzero eigenvalue of ${\cal M}$ (particles masses are set to unity).

In Fig.~\ref{MVF_fig}a the sample-to-sample means $\langle \omega_{\rm min} \rangle$, rescaled by $2\pi\sqrt{\mu/\rho}/a_0$, are plotted vs.~the system size $L$. Here $a_0$ is a microscopic lengthscale that characterizes the pairwise potential, $\mu$ is the athermal shear modulus~\cite{athermal_elasticity} and $\rho\!\equiv\!N/V$ is the density with $V\!=\!L^\dbar$. We find that for all models considered and systems of size $L\!\lesssim\!L_D\!\approx\!60$ (in our microscopic units) $\langle\omega_{\rm min}\rangle\!\sim\!L^{-3/5}$. Equation~(\ref{meanMVF}) then suggests that $\theta\!=\!4$.

In Figs.~\ref{MVF_fig}b-c we plot the sample-to-sample distributions of MVFs $P(\omega_{\rm min})$ measured for the 3DIPL system. Panel (b) shows the raw distributions, while in panel (c) the same distributions are shown in terms of the rescaled variable $\omega_{\rm min}L^{3/5}$, following Eq.~(\ref{mvf_dist}) with $\theta\!=\!4$. The rescaling assuming Weibullian statistics leads to an essentially perfect collapse of the distributions. The continuous magenta line represents the Weibull distribution $W(y)\!\propto\!y^4e^{-(y/y_0)^5}$, with $y_0\!\approx\!4$. The quality of this collapse constitutes additional strong evidence for the robustness of the $D(\omega)\!\sim\!\omega^4$ law.

\begin{figure}[!ht]
\centering
\includegraphics[width = 0.495\textwidth]{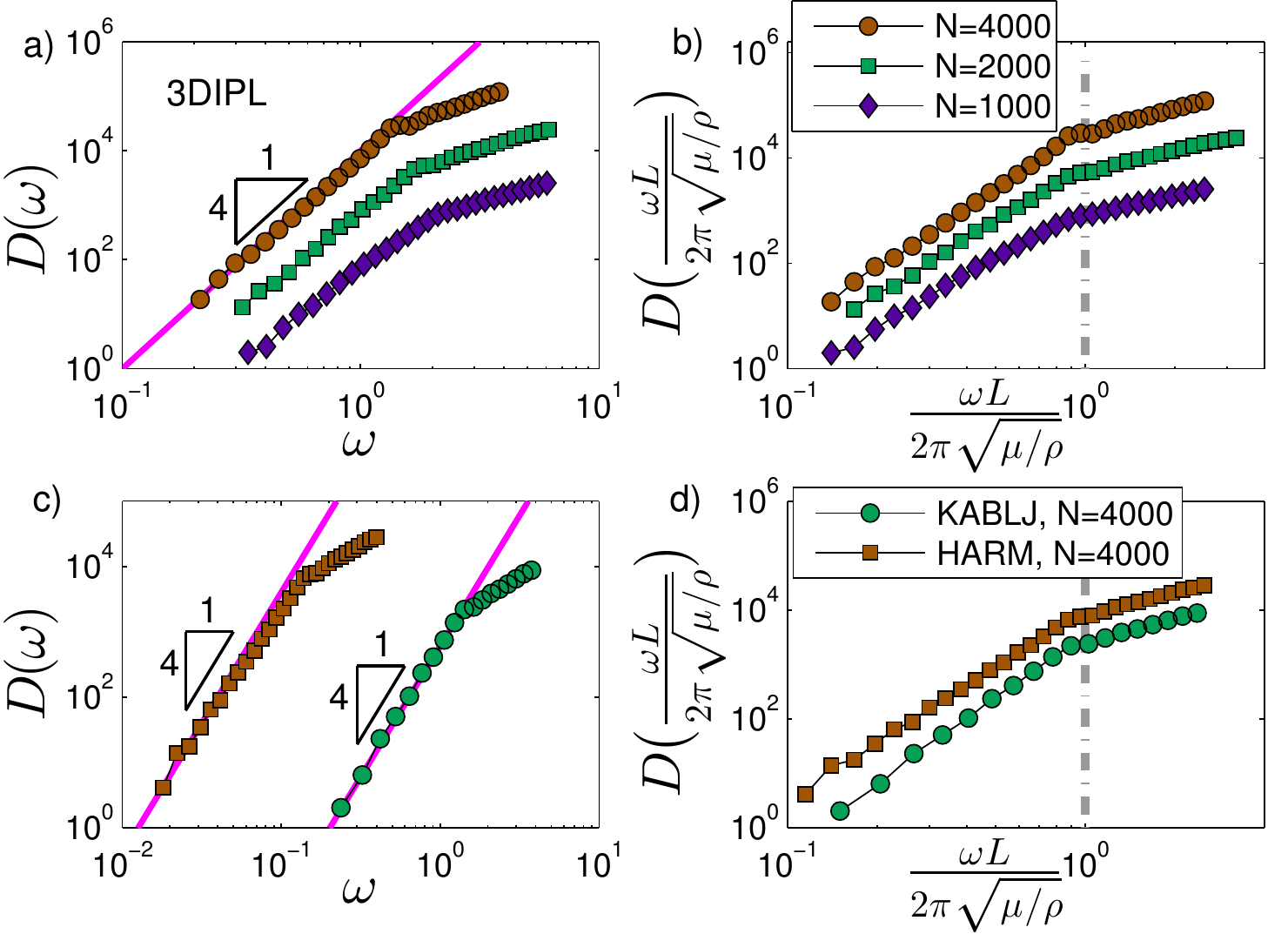}
\caption{\footnotesize Left column: density of vibrational modes $D(\omega)$ measured in (a) the 3DIPL system and (c) the KABLJ and HARM systems. The continuous magenta lines all represent the scaling $D(\omega)\!\sim\!\omega^4$. Right column: same distributions, plotted vs.~the rescaled frequencies $\omega L/(2\pi\sqrt{\mu/\rho})$. The vertical dashed lines represent the lowest Goldstone mode frequency expectation. Distributions were shifted vertically for visibility.}
\label{dos_fig}
\end{figure}

These results suggest that in systems with $L\!\ll\!L_D\!\approx\!60$, a low-frequency tail of the form $D(\omega)\!\sim\!\omega^4$ should be directly observable. Guided by these results, the low-frequency tails of $D(\omega)$ were calculated and plotted in Fig.~\ref{dos_fig} for all the aforementioned glass-forming models and various system sizes. The left columns display the raw distributions, while in the right column we plotted the same distributions as a function of the frequencies rescaled by the lowest Goldstone mode frequency $\omega L/(2\pi\sqrt{\mu/\rho})$. The magenta (solid) lines correspond to $D(\omega)\!\sim\!\omega^4$, which appears to be followed by the data for all models, up to the vicinity of the lowest Goldstone mode frequency (indicated by the dash-dotted line).

\begin{figure}[!ht]
\centering
\includegraphics[width = 0.48\textwidth]{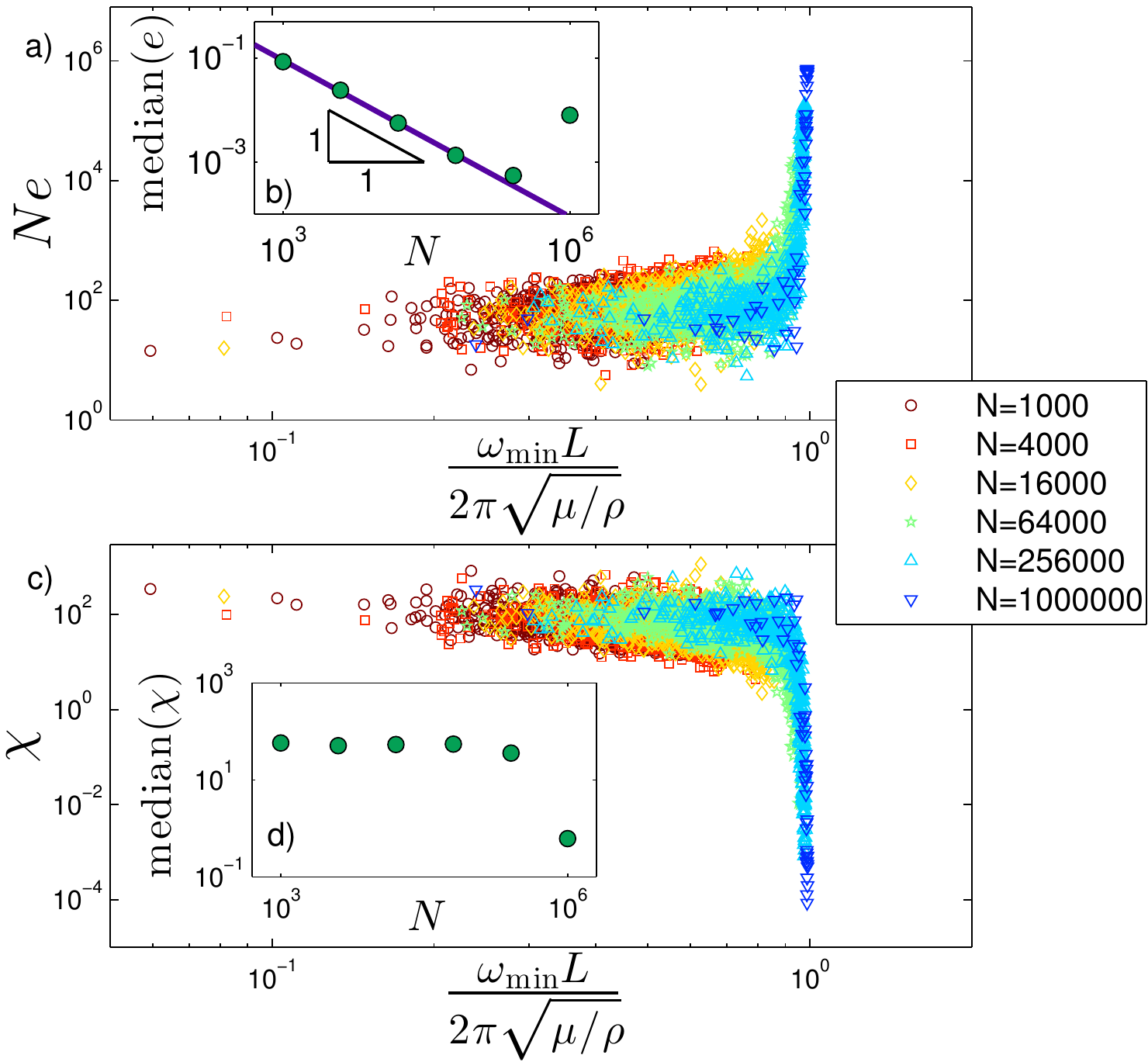}
\caption{\footnotesize Scatter plots of the product $Ne$ (a) and anharmonicity $\chi$ (b) vs.~the rescaled MVF.
Median participation ratio $e$ and anharmonicity $\chi$ vs.~system size are shown in the insets of panels (a) and (b), respectively.}
\label{chi_fig}
\end{figure}

\emph{Localization and anharmonicity.--} Once the $\omega^4$ scaling is established, we study next the localization properties and anharmonicity of the lowest frequency modes. We first consider the participation ratio $e\!\equiv\!\big(N\sum_i (\hat{\Psi}_i\cdot\hat{\Psi}_i)^2\big)^{-1}$ of the lowest frequency modes $\hat{\Psi}$, which is an indicator of the degree of their spatial localization. Fig.~\ref{chi_fig}a shows a scatter plot of the products $Ne$ vs.~the rescaled MVF. There appears to be no clear correlation between the localization degree of the softest modes and their frequencies for $\omega_{\rm min}\!<\!2\pi\sqrt{\mu/\rho}/L$. The inset shows the median of $e$ vs.~system size $N$, revealing a clear $e\!\sim\!N^{-1}$ scaling for $N\!\le\!64000\!<\!\rho L_D^3$. This indicates that the lowest frequency modes are quasi-localized, supporting similar conclusions by Schober and coworkers \cite{Schober_Laird_numerics_PRL,Schober_Laird_numerics_PRB,Schober_Oligschleger_numerics_PRB}. 

In Fig.~\ref{chi_fig}c a scatter plot of the quartic anharmonicity $\chi\!\equiv\!\frac{\partial^4U}{\partial\vec{x}_i\partial\vec{x}_j\partial\vec{x}_k\partial\vec{x}_\ell}\!\quadCdot\!\hat{\Psi}_i\hat{\Psi}_j\hat{\Psi}_k\hat{\Psi}_\ell$ associated with the lowest frequency modes vs.~the rescaled MVF is presented. We observe that the anharmonicity of the softest modes is also not correlated with their frequencies, as long as the latter are smaller than the lowest Goldstone mode frequency. In addition, the anharmonicity is $N$-independent for systems with $L\!<\!L_D$ (see inset).

\begin{figure}[!ht]
\centering
\includegraphics[width = 0.50\textwidth]{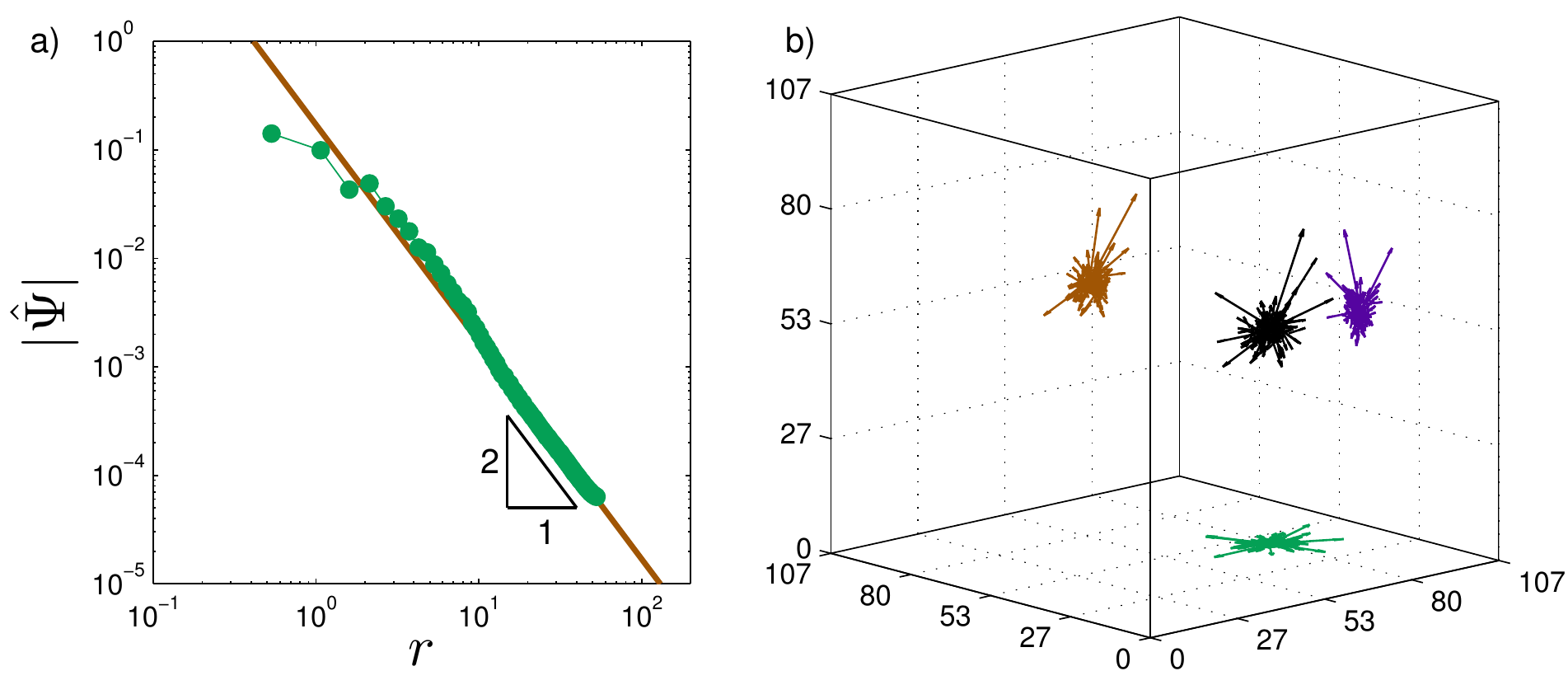}
\caption{\footnotesize  (a) Spatial decay profile (see Appendix  for exact definition) of the lowest frequency mode amongst our entire ensemble of minimal frequency modes with $N\!=\!10^6$. (b) The ensemble-lowest frequency mode (only components larger than a tenth of the mode's maximal component are shown). On the $x$-$y$, $x$-$z$ and $y$-$z$ planes the respective projections of the mode are shown, allowing for a visual estimation of its spatial scale.}
\label{decay_profiles}
\end{figure}

To further explore the localization properties, we show in Fig.~\ref{decay_profiles}a that the spatial profile of the lowest frequency mode (amongst our entire ensemble of minimal frequency modes in systems with $N\!=\!10^6$) decays as $r^{-2}$ for $r\!\gtrsim\!\xi_g\!\approx\!10$, see Appendix for definitions and procedures. This same decay profile was found for destabilizing modes at the onset of plastic instabilities in externally deformed athermal glasses~\cite{luka,lemaitre2006_avalanches}. In Fig.~\ref{decay_profiles}b we show the ensemble-lowest mode itself, demonstrating a core size consistent with $\xi_g\!\approx\!10$, as estimated from the decay profile. We identify $\xi_g$ as the localization length of quasi-localized soft modes.

\emph{Preparation protocols and lengthscales.--} Recent experiments suggest that glasses created by careful vapor deposition techniques~\cite{deposition} are free of low-frequency glassy modes, as indicated by the crystalline-like temperature dependence of their specific heat~\cite{ultrastable_perez_pnas} and by their suppressed $\beta$-relaxation~\cite{ultrastable_ediger}. It is therefore important to test whether the observed $\omega^4$ law and the Weibullian statistics of MVFs are affected by the preparation protocol. 

To this aim, in addition to the rapidly quenched glasses discussed up to now, we also prepared an ensemble of glassy samples that were slowly quenched through the computer glass transition, see Appendix for further details. In Fig.~\ref{quench_rate_fig}a we plot $\langle \omega_{\rm min} \rangle$, rescaled by $2\pi\sqrt{\mu/\rho}/a_0$, vs.~$L$ for the rapidly and slowly quenched ensembles. It is observed that the slower quenched glasses still exhibit $\langle \omega_{\rm min}\rangle\!\sim\!L^{-3/5}$ scaling, indicating the robustness of the $\omega^4$ law to different preparation protocols (see further discussion in the Appendix). Note that the rescaled $\langle \omega_{\rm min}\rangle$(L) does not collapse onto a single curve for the two ensembles, which implies that the preparation protocol dependence of $\langle \omega_{\rm min}\rangle$ and $\mu$ (in fact $\sqrt{\mu}$) is different. We address this point next.

In Figs.~\ref{quench_rate_fig}b-c the sample-to-sample distributions of $\langle \omega_{\rm min}\rangle$ and $\mu$ are plotted, respectively. Both distributions exhibit stiffening as the cooling rate decreases, though $\langle \omega_{\rm min}\rangle$ stiffens significantly more strongly (the mean shifts to a higher frequency by roughly 25\%) than $\mu$ (the mean shifts by roughly 10\%), which is consistent with the non-collapse observed in Fig.~\ref{quench_rate_fig}a. Finally, in Fig.~\ref{quench_rate_fig}d the sample-to-sample distribution of the participation ratio $e$ is plotted, indicating that the glassy soft modes become more localized when the samples are cooled more slowly. This implies that the localization length $\xi_g$ decreases with decreasing cooling rate.

\begin{figure}[!ht]
\centering
\includegraphics[width = 0.485\textwidth]{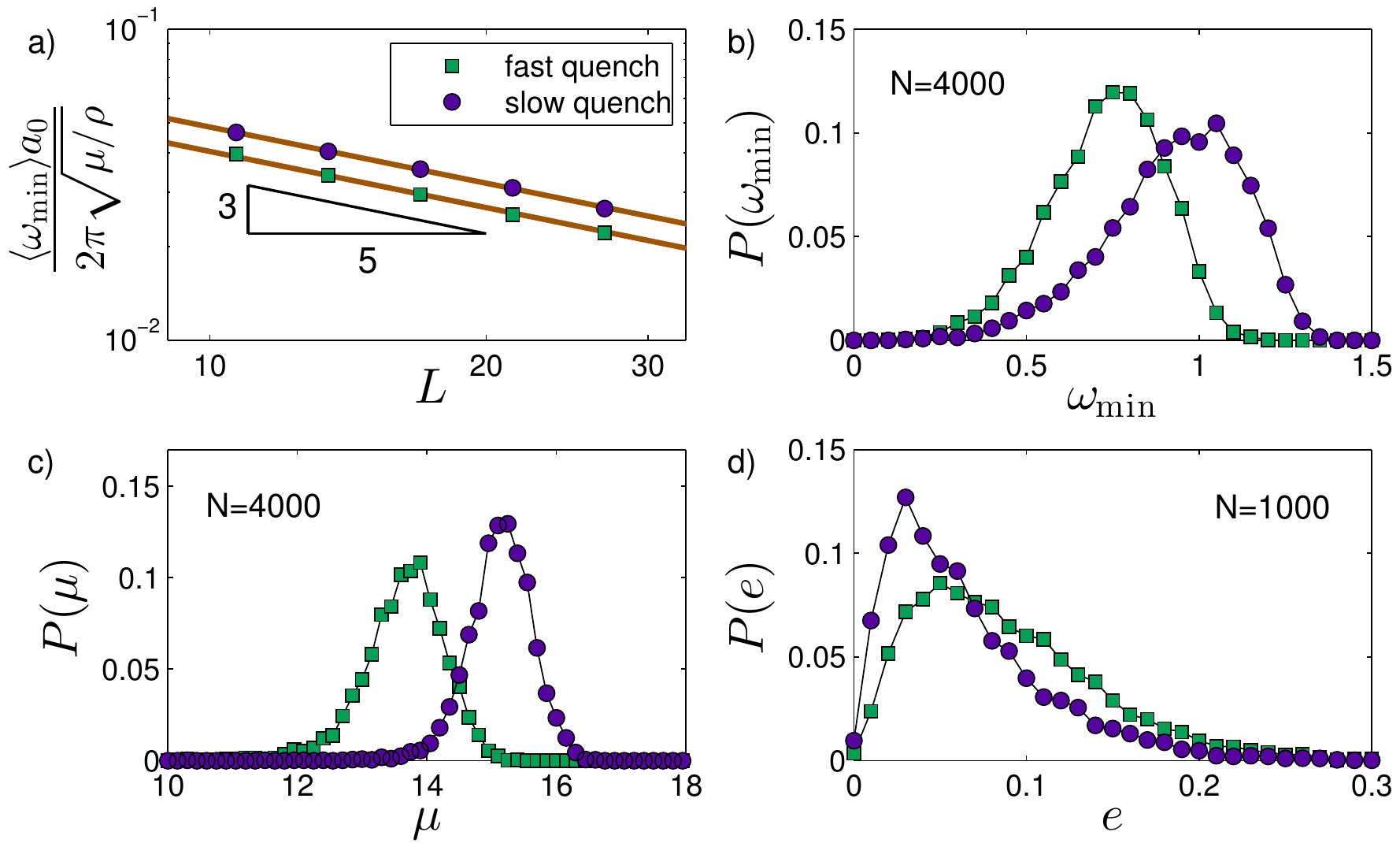}
\caption{\footnotesize  (a) The mean rescaled MVFs depend on system size as $L^{-3/5}$, both for rapidly and slowly quenched samples, suggesting the robustness of the $\omega^4$ law to different preparation protocols. (b)-(d) Distributions of (b) MVFs $\omega_{\rm min}$, (c) athermal shear modulus $\mu$, and (d) participation ratio $e$ measured in our slowly and rapidly quenched solids as indicated by the legend. See text for discussion.}
\label{quench_rate_fig}
\end{figure}

How should one interpret these preparation protocol dependencies and their relation to glassy lengthscales? To address this question, we rewrite Eq.~\eqref{meanMVF} in dimensional form as $\langle\omega_{\rm min}\rangle\!\sim\!\omega_g\left(L/\xi_s\right)^{-\dbar/5}$. Here $\omega_g$ is a ``glassy modes'' characteristic frequency scale, which must appear in a parent distribution $P_0(\omega)\!\sim\!\omega_g^{-1}(\omega/\omega_g)^4$ associated with Weibullian statistics, and $\xi_s$ is a ``site length'' that implies that $\langle \omega_{\rm min}\rangle$ of each sample is the softest mode amongst $(L/\xi_s)^\dbar$ candidates. $\omega_g$ is generally preparation protocol dependent and is expected to stiffen with decreasing cooling rates. While this behavior rationalizes the observed trend in $\langle \omega_{\rm min}\rangle$, we cannot rule out the possibility that $\xi_s$ is also protocol dependent, which implies that the stiffening of $\langle\omega_{\rm min}\rangle$ may not be wholly explained by the stiffening of $\omega_g$ (see related discussion in~\cite{wyart_vibrational_entropy}). At this point, however, we are unable to disentangle the preparation protocol dependencies of $\omega_g$ and $\xi_s$.

Up to now $3$ glassy lengthscales were mentioned: the crossover length $L_D$, the localization length $\xi_g$ and the site length $\xi_s$. We briefly note that the crossover length $L_D$ is determined by the system size at which the lowest glassy mode frequency is of the order of the lowest Goldstone frequency, i.e.~$\omega_g \left(L_D/\xi_s\right)^{-3/5}\!\sim\! L_D^{-1}\sqrt{\mu/\rho}$. This leads to $L_D\!\sim\!\xi_{BP}\left(\xi_{BP}/\xi_s\right)^{3/2}$,
where we identified yet another lengthscale $\xi_{BP}\!\equiv\!\omega_g^{-1}\sqrt{\mu/\rho}$, the ``boson peak'' lengthscale, closely related to the one introduced in e.g.~\cite{sokolov_boson_peak_scale}. Understanding the relations between these lengthscales and their dependence on the preparation protocol is an important task to be further addressed in a separate report.

\emph{Concluding remarks and prospects.--} In this Letter we showed that the distribution of low-frequency vibrational {\em glassy modes} in several 3D models of structural glasses follows a $\omega^4$ law. This scaling is observable by carefully tuning the system size such that Goldstone modes are suppressed. In addition, the sample-to-sample statistics of MVFs is shown to be Weibullian, with scaling exponents that are fully consistent with the $\omega^4$ law.

Our results also establish the existence of a preparation protocol dependent localization length that characterizes soft glassy modes, and that the anharmonicity associated with these modes is frequency and system size independent. These are two of the key assumptions made in the ``Soft Potential Model" \cite{soft_potential_model} that predicts the $\omega^4$ law for soft glassy modes. It is desirable to extend our numerical analysis towards the validation of the more recent ``reconstruction picture" \cite{Gurevich2003,Gurevich2007}, in which interactions between different localized excitations and anharmonicity give rise to the $\omega^4$ law for soft glassy modes.

We have only reported here results for 3D systems. Preliminary results indicate that the $\omega^4$ law persists in the density of states of 2D glasses of sizes $L\!\ll\! L_D(\dbar = 2)$. However, we find that the Weibullian statistics of MVF breaks down in 2D, as do the quasi-localization of lowest frequency modes and $N$-invariance of their associated anharmonicity. These issues will be addressed is a separate, broader report.

\textit{Acknowledgments.--} We thank Yohai Bar-Sinai and Matthieu Wyart for fruitful discussions. We acknowledge Smarajit Karmakar for first proposing to study the sample-to-sample statistics of the lowest frequency modes. E.L.~acknowledges support from the Amsterdam Academic Alliance fellowship. G.D.~acknowledges support from FONDECYT Grant No.~1150463. E.B.~acknowledges support from the Israel Science Foundation (Grant No.~712/12), the Harold Perlman Family Foundation and the William Z. and Eda Bess Novick Young Scientist Fund. 

\vskip 0.17cm

E.L.~designed and performed the research, E.L., G.D.~and E.B.~discussed the results and E.L.~and E.B.~wrote the Letter.

\appendix
\section{Appendix}
In this Appendix we $(i)$ describe the models and numerical methods used to obtain the results presented in the manuscript, and $(ii)$ provide a short discussion regarding the statistics of minimal vibrational frequencies in our ensemble of slowly quenched solids.

\subsection*{Numerical methods}
We employ three popular glass forming models in three dimensions: 
\begin{enumerate}
\item{(HARM) A binary mixture of `large' and `small' soft spheres of equal mass $m$ interacting via a one-sided harmonic radially-symmetric pairwise potential of the form
\begin{equation}
\varphi_{\mbox{\tiny HARM}}(r_{ij}) =  \left\{ \begin{array}{ccc}\sFrac{1}{2}k\big(r_{ij} - (\sigma_i + \sigma_j) \big)^2&,& r_{ij} \le \sigma_i + \sigma_j\\0&,&r_{ij} >  \sigma_i + \sigma_j\end{array} \right.\,,
\end{equation}
where $r_{ij}$ is the distance between the centers of the $i^{\mbox{\tiny th}}$ and $j^{\mbox{\tiny th}}$ spheres, $k$ is a stiffness constant, and $\sigma_i$ denotes the radius of the $i^{\mbox{\tiny th}}$ sphere. We used a 50:50 binary mixture, where half the particles have a radius of $0.5a_0$ and the other half of $0.7a_0$. The microscopic unit of length $a_0$ was chosen as the diameter of the small particles. $m$ denotes the units of mass, energies are expressed in units of $ka_0^2$, temperatures in units of $ka_0^2/k_B$ (with $k_B$ being the Boltzmann constant), pressure in units of $k/a_0$ and time in units of $\sqrt{m}/(ka_0)$. We prepared packings with a pressure $p\!=\!10^{-1}$ and zero temperature by first equilibrating systems at the density $N/V \!=\! 1.0$ and temperature $T\!=\!0.5$ for $100.0$ microscopic time units, followed by an energy minimization using a combination of the FIRE algorithm \cite{fire} coupled to a Berensden barostat \cite{berendsen}.
}

\item{(KABLJ) The canonical Kob-Andersen binary Lennard-Jones system \cite{kablj} is a binary mixture of 80\% type A particles and 20\% type B particles, interacting via the following radially-symmetric pairwise potential
\begin{widetext}
\begin{equation}
\varphi_{\mbox{\tiny LJ}}(r_{ij}) = \left\{ \begin{array}{ccc}\varepsilon_{ij}\left[ \left( \sFrac{\sigma_{ij}}{r_{ij}} \right)^{12} - \left( \sFrac{\sigma_{ij}}{r_{ij}} \right)^6 + c_6\left(\sFrac{r_{ij}}{\sigma_{ij}}\right)^6 + c_4\left(\sFrac{r_{ij}}{\sigma_{ij}}\right)^4 + c_2\left(\sFrac{r_{ij}}{\sigma_{ij}}\right)^2 + c_0\right]&,&\sFrac{r_{ij}}{\sigma_{ij}}\le x_c\\0&,&\sFrac{r_{ij}}{\sigma_{ij}}> x_c\end{array} \right.\,.
\end{equation}
\end{widetext}
Energies are expressed in terms of $\varepsilon_{AA}$, then $\varepsilon_{AB} \!=\! 1.5$ and $\varepsilon_{BB} \!=\! 0.5$. The interaction length parameters are expressed in terms of $a_0 \!\equiv\! \sigma_{AA}$, then $\sigma_{AB} \!=\! 0.8$ and $\sigma_{BB} \!=\! 0.88$. $x_c \!=\! 2.5$ is the dimensionless distance for which $\varphi_{\mbox{\tiny LJ}}$ vanishes continuously up to $3$ derivatives, and the density was set at $N/V\!=\!1.2$. Temperature is expressed in terms of $\varepsilon_{AA}/k_B$ with $k_B$ the Boltzmann constant. Time is expressed in terms of $\sqrt{ma_0^2}/\varepsilon_{AA}$, with $m$ denoting the microscopic units of mass. With this parameter set the system experienced a computer glass transition at $T_g\!\approx\! 0.45$. Solids were prepared by equilibrating systems at $T\!=\!1.0$ for 50.0 time units, \hl{followed by an anealing run of 50.0 time units at $T\!=\!0.45$} \cite{footnote}. Finally, a rapid quench to $T\!=\!0$ was performed by means of a conventional conjugate gradient algorithm.
}

\begin{figure*}[!ht]
\centering
\includegraphics[scale = 0.6]{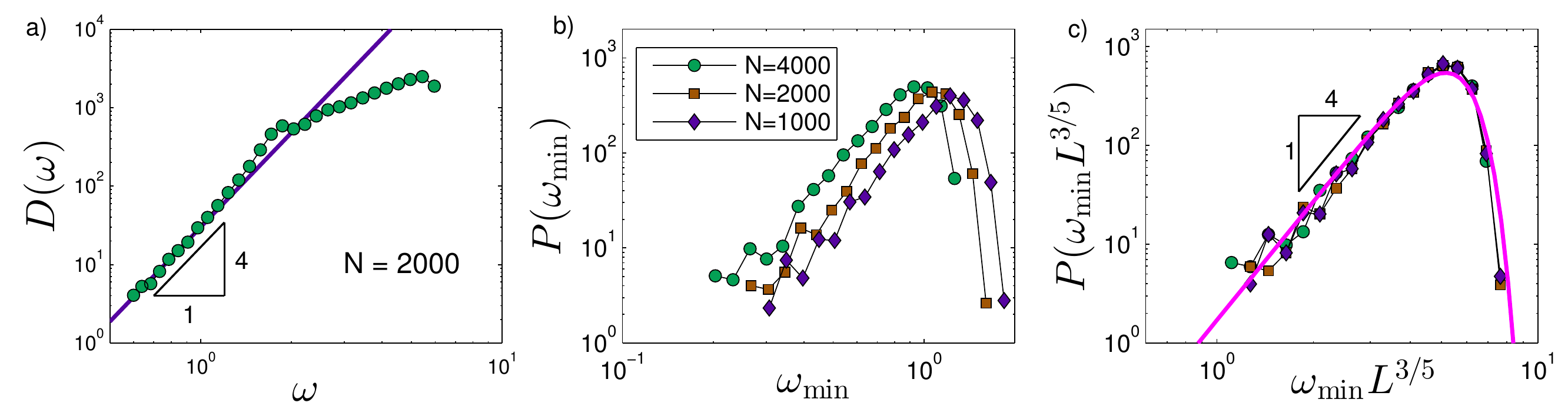}
\caption{\footnotesize Statistics of low-frequency vibrational modes in slowly quenched glasses. (a) Low frequency regime of the density of states of the 3DIPL system with $N\!=\!2000$. (b) Distributions of $\omega_{\rm min}$ for system sizes as indicated by the legend. (c) Same distributions as in panel (b), plotted as a function of the rescaled variables $\omega_{\rm min}L^{3/5}$. The continuous magenta line represents the Weibull distribution, see text for details.}
\label{slow_quench_min_freq_dist}
\end{figure*}

\item{(3DIPL) A 50:50 binary mixture of `large' and `small' particles of equal mass $m$, interacting via radially-symmetric purely repulsive inverse power-law pairwise potentials, that follow
\begin{equation}
\varphi_{\mbox{\tiny IPL}}(r_{ij}) = \left\{ \begin{array}{ccc}\varepsilon\left[ \left( \sFrac{\sigma_{ij}}{r_{ij}} \right)^n + \sum\limits_{\ell=0}^q c_{2\ell}\left(\sFrac{r_{ij}}{\sigma_{ij}}\right)^{2\ell}\right]&,&\sFrac{r_{ij}}{\sigma_{ij}}\le x_c\\0&,&\sFrac{r_{ij}}{\sigma_{ij}}> x_c\end{array} \right.,
\end{equation}
where $r_{ij}$ is the distance between the $i^{\mbox{\tiny th}}$ and $j^{\mbox{\tiny th}}$ particles, $\varepsilon$ is an energy scale, and $x_c$ is the dimensionless distance for which $\varphi_{\mbox{\tiny IPL}}$ vanishes continuously up to $q$ derivatives. Distances are measured in terms of the interaction lengthscale $a_0$ between two `small' particles, and the rest are chosen to be $\sigma_{ij} \!=\! 1.18a_0$ for one `small' and one `large' particle, and $\sigma_{ij} \!=\! 1.4a_0$ for two `large' particles. The coefficients $c_{2\ell}$ are given by
\begin{equation}
c_{2\ell} = \frac{(-1)^{\ell+1}}{(2q-2\ell)!!(2\ell)!!}\frac{(n+2q)!!}{(n-2)!!(n+2\ell)}x_c^{-(n+2\ell)}\,.
\end{equation}
We chose the parameters $x_c = 1.48, n=10$, and $q=3$. The density was set to be $N/V\!=\!0.82a_0^{-3}$. Temperatures are expressed in terms of $\varepsilon/k_B$ with $k_B$ the Boltzmann constant, and time in terms of $\sqrt{ma_0^2}/\varepsilon$, with $m$ denoting the microscopic units of mass. This system undergoes a computer glass transition at $T_g\!\approx\!0.5$. Solids were created by first equilibrating system at $T\!=\!1.0$, \hl{followed by an annealing run of 100.0 time units at $T\!=\!0.5$} \cite{footnote}. Finally, a rapid quench to zero temperature was carried out by means of a conventional conjugate gradient algorithm. We have also created an ensemble of slowly quenched solids (see data and discussion in main text), cooled at a rate of $10^{-5}$ through the glass transition. 

}
\end{enumerate}

The stopping condition for our minimizations was set as follows; we calculate a characteristic interaction force scale $\bar{f}\!\equiv\!\left( \sum_\alpha f_\alpha^2/N\right)^{1/2}$ and a characteristic net force scale $\bar{F}\!\equiv\! \left( \sum_i |\vec{F}_i|^2/N\right)^{1/2}$, where $\alpha$ labels a pair of interacting particles, $f_\alpha\! \equiv\! -\frac{\partial \varphi}{\partial r_\alpha}$ is the force exerted between the $\alpha^{\mbox{\tiny th}}$ pair, $\vec{F}_i\!\equiv\! -\frac{\partial U}{\partial\vec{x}_i}$ is the net force experienced by the $i^{\mbox{\tiny th}}$ particle, and $N$ is the number of particles in the sample. We then terminate the minimization algorithm once the ratio $\bar{F}/\bar{f}$ drops below $10^{-10}$. 

Normal modes were calculated both using Matlab \cite{matlab}, and following the methods presented in \cite{micromechanics2016}. We have validated by comparison of the two methods and resorting to 128-bit precision that our analysis does not suffer from numerical inaccuracies.

We finally explain here how the spatial decay profile as shown in Fig.~4 of the main text was calculated. Given a mode $\hat{\Psi}$, we identify the mode's core as explained in \cite{luka}. We then calculate the median of the square of $\hat{\Psi}$'s components over a thin spherical shell, with thickness on the order of $a_0$, and of radius $r$ away from the mode's core. The decay profiles are defined as the square root of these medians.  

\subsection*{Statistics of low frequency modes in slowly quenched samples}
In Fig.~\ref{slow_quench_min_freq_dist}a we show the direct calculation of the density of states $D(\omega)$ for the slowly-quenched 3DIPL systems with $N\!=\!2000$. In panel (b) we show the distributions $P(\omega_{\rm min})$ of minimal vibrational frequencies calculated for the ensemble of slowly-quenched solids. In panel (c) we plot the same distributions, but this time as a function of the rescaled minimal frequencies $\omega_{\rm min}L^{3/5}$. The continuous magenta lines correspond to the Weibull distribution $W(y)\!\propto\!y^4e^{-(y/y_0)^5}$, with $y_0\!\approx\!5.4$. The analysis is restricted to small systems due to poor statistics for larger systems. It is clear that the rescaling of the minimal vibrational frequencies by $L^{-3/5}$ leads to a very good collapse of the distributions. All aformationed data indicates that the $\omega^4$ law persists under a careful quench of structural glasses.

\bibliography{references_lerner}

\begin{thebibliography}{48}
\bibitem [{foo()}]{footnote}%
  \BibitemOpen
  \href@noop {} {\bibinfo  {journal} {These annealing runs were carried out to
  prevent numerical instabilities that would occur from time to time when
  attempting to minimize the energy of high temperature configurations using a
  nonlinear conjugate gradient algorithm, see also \cite{modes_prl_erratum}}\
  }\BibitemShut {NoStop}%
\end{thebibliography}

\end{document}